\documentclass{aa}  

\usepackage{graphicx}
\usepackage{txfonts}
\usepackage{supertabular}
\usepackage{amsmath}

\newcommand{\oh}{OH\,231.8+4.2\,}

\newcommand{\kms}{km~s$^{-1}$}
\begin{document}

   \title{On the Nature of the Protoplanetary Nebula OH\,231.8+4.2}


    \author{Andreas Brunthaler
          \inst{1}
          \and
          Yoon Kyung Choi \inst{1} \and Karl M. Menten \inst{1} \and Mark J. Reid \inst{2} 
          }
 
      \institute{Max-Planck-Institut f$\ddot{\textrm u}$r Radioastronomie, 
              Auf dem H$\ddot{\textrm u}$gel 69, 53121 Bonn, Germany \\
              \email{[brunthal;ykchoi;kmenten]@mpifr-bonn.mpg.de}
              \and
	              Center for Astrophysics $\vert$ Harvard \& 
               Smithsonian,
              60 Garden Street, Cambridge, MA 02138, USA \\
             \email{reid@cfa.harvard.edu}
              }

   \date{Received; accepted }

 
  \abstract
  {}
  {\oh, also known as the Rotten Egg or Calabash nebula is a protoplanetary nebula which is seen in the direction of the open cluster M\,46. While an association of the nebula with the cluster has been suggested in the past, this has been never confirmed. Here, we present accurate trigonometric parallax and proper motion measurements using VLBI observations of masers in the nebula and Gaia DR3 data for the cluster.}
   {We observed 22 GHz H$_2$O and 43 GHz SiO masers around OH\,231.8+4.2 using the Very Long Baseline Array at 4 epochs over 1 year. We also calculated the parallax and proper motion of the open star cluster M\,46 using Gaia DR3 data.}
   {Based on astrometric monitoring for 1 year, we measured an annual parallax of OH\,231.8+4.2 to be 0.65 $\pm$ 0.01 mas (stat.) $\pm$ 0.02 mas (syst.), corresponding to a distance of 1.54 $\pm$ 0.05 kpc. This agrees well with the parallax of M\,46 from the Gaia DR3 data, which is 0.639 $\pm$ 0.001 mas (stat.) $\pm$0.010 mas (syst.). We re-estimated the luminosity of OH\,231.8+4.2 to be 1.4 $\times$ 10$^4$ L$_{\odot}$. However, there is a ~15\kms velocity difference between \oh and M\,46 which could be caused by a past merger event.}
   {}

   \maketitle
%

\section{Introduction}

A protoplanetary nebula (PPN) is in a transition phase from an asymptotic giant branch (AGB) star to a planetary nebula (PN). 
The transition lasts only a short time (thousands of years) in the evolution of low mass stars. While most AGB stars have approximately spherically symmetric circumstellar envelopes, PPNe and PNe show strong asymmetries, frequently collimated bipolar winds. H$_2$O and SiO maser emission may arise in regions very close to the photosphere of the central object of a PPN, which still has AGB stellar characteristics, and in which hydrogen shell burning has not yet raised the central star's effective temperature to prohibitively high levels. Since these masers form in the inner circumstellar envelope \citep[see, e.g.,][]{Reid1990, Vlemmings, Reid2007, Imai}, they are good probes of the formation of bipolarity. 

OH\,231.8+4.2 (or OH\,0739$-$14) has been classified as a PPN, showing OH, H$_2$O and SiO maser emission. It was first discovered by its peculiar 18 cm OH maser emission, which covered an uncommonly wide velocity range in (only) the 1667 MHz hyperfine structure line \citep{Turner1971}. OH\,231.8+4.2 has been named the Calabash nebula, given its morphology of a shocked bipolar outflow \citep[see, for example][]{Reipurth1987}]
and the Rotten Egg Nebula, because of the prevalence of 
sulphur bearing molecules, such as SO, SO$_2$ and H$_2$S that highlight its rich and (for an evolved star) peculiar chemical composition \citep{Morris1987}.

According to \citet{Sanchez}, the central source of OH\,238.8+4.2 is a binary system, composed of an M9-10 III Mira variable (which is, an AGB star) and an A0 main sequence companion. It shows a highly collimated bipolar outflow in CO emission ($J = 1 - 0$ and $2 - 1$ lines) \citep{Alcolea01}, with an dynamic age of about 800 years \citep{Bu02}. It has been suggested that gas accreting from the Mira star to a companion in a close interaction formed an accretion disk powering the outflows \citep{Soker12}. Based on this scenario \cite{Staff2016} performed  hydrodynamical simulations  that showed that this interaction would have lead to a merger of the two stars, indicating that the system was initially a triple system.

The central Mira star in OH\,231.8+4.2, also known as QX Pup, is variable with estimated periods of 648 \citep{Feast83} and 708 days \citep{Kastner92} and a luminosity between 10$^4$ to 2 $\times$ 10$^4$ L$_{\odot}$ \citep{Kastner98}. 

Previous VLBI observations \citep{Desmurs} determined
the positions of the H$_2$O ($6_{1,6}-5_{2,3}$) and SiO ($v$ = 2, $J$ = 1 -- 0) masers.
The H$_2$O maser clumps were detected in two regions
separated by 60 mas in the north-south direction, roughly along the axis of the much 
larger scale bipolar outflow; also the 
blue- and red-shifted masers are distributed in the same sense as on larger scales. 
In Miras, SiO masers, are generally located within $\approx5$ stellar radii.  Therefore they  indicate the position of the Mira component of the binary system. 
In OH\,231.8+4.2 the SiO maser emission is found between the two H$_2$O maser regions and has a distribution that
is elongated perpendicular to the nebular axis, 
suggesting the presence of an equatorial torus or disk around the central star. 


An association of OH\,231.8+4.2 
with the open star cluster M\,46 has been suggested by \citet{Jura}, 
whose distance has been estimated to be 1.3$\pm$0.4 kpc \citep{Kastner92}. More recently, estimates for the distance to M\,46 using Gaia DR2 data range from 1.64 kpc  \citep{Cantat18} to 1.67 kpc \citep{Babusiaux18}. The age of the cluster is about 320 Myr \citep{Tarricq}. Note that the PN NGC\,2438 is also seen in the direction of M\,46. But it has been shown that this PN is a foreground object \citep{Kiss2008}.
 
While OH\,231.8+4.2 is also listed in the first VERA catalog \citep{Hirota2020} with a distance of 1.64$\pm$0.08 kpc, the paper which is cited for this value has not been published yet. Since the central star is heavily obscured at optical wavelength, a direct parallax measurement of OH\,231.8+4.2 with Gaia is not possible. However, a reliable distance is crucial to study the properties of the nebula. Trigonometric parallaxes of circumstellar maser sources (H$_2$O and/or SiO) in late-type stars have been measured with VLBI phase-referencing with accuracies of $\pm40$ to 80 $\mu$as \citep{Choi, Tafoya, Zhang12a, Zhang12b, Nakagawa2016}. 

In this paper, we present an accurate trigonometric parallax measurement of OH\,231.8+4.2.
In Section 2, we describe the observations and data reduction, and in Section 3 we present our results. 
In Section 4, we give an update on the distance to the open cluster M\,46 Gaia DR3. Finally, we summarize the most important results in Section 5.

\section{VLBA Observations and data reduction}

\begin{table*}[t]
	\centering  
	\caption{Source information.}             
	\label{table:survey}      
        
	\begin{tabular}{l l l r r r r c}     
	\hline\hline       
	Source &  R.A. (J2000) & Decl. (J2000)     & $\theta_{\rm sep}$ & P.A. & $I_{peak}$ (22 GHz) & $I_{peak}$ (43 GHz) & Ref. \\ 
              & (h \ m \ s)    & ($^\circ$ \ ' \ ") & ($^\circ$) & ($^\circ$) & (mJy beam$^{-1}$) & (mJy beam$^{-1}$) & \\
	\hline
	OH\,231.8+4.2 &  07 \ 42 \ 16.91994 & --14 \ 42 \ 50.0574 & -- & -- & &  & \\
	J0746--1555 &  07 \ 46 \ 18.2360 & --15 \ 55 \ 34.746 & 1.55 & 141 &  138.64  & \ 76.45 & 1 \\
	J0748--1639 &  07 \ 48 \ 03.0838 & --16 \ 39 \ 50.255 & 2.39 & 145 &  296.77  &  168.47 & 3 \\
	J0748--1650 &  07 \ 48 \ 48.6786 & --16 \ 50 \ 27.401 & 2.64 & 144 & \ 71.56  &    --    & 1 \\
	J0735--1735 &  07 \ 35 \ 45.8125 & --17 \ 35 \ 48.502 & 3.28 & --151 &  -- & -- & 4 \\
	J0729--1320 &  07 \ 29 \ 17.8177 & --13 \ 20 \ 02.272 & 3.44 & --66 &  \ 60.61  & --  & 2 \\
	J0756--1542 &  07 \ 56 \ 50.6990 & --15 \ 42 \ 05.436 & 3.47 & 107 & \ 163.95  & 172.94 & 3 \\ 
	\hline               
	\end{tabular}
\tablefoot{Absolute positions are accurate to better than 1 mas in each coordinate. 
$\theta_{\rm sep}$ and P.A. are angular separations and position angles east of north 
of the background sources relative to OH\,231.8+4.2 in Columns 4 and 5.
Peak intensities in Columns 6 and 7 were obtained in our calibrator survey. Dashes indicate a non-detection.
References. --
(1) VCS2 \citep{VCS2}; (2) VCS3 \citep{VCS3}; (3) VCS4 \citep{VCS4}; 
(4) GSFC from the ICRF2 catalog \citep{Ma2009}.}

\end{table*}

\begin{table*}
\begin{center}
         \caption[]{Observation information}
         \label{table:obs}   
         \begin{tabular}{c l c l c c}
         \hline\hline
         Epoch & Code & Date & Participating antennas & Restoring beam & rms noise \\
               &      & &                        & (mas $\times$ mas, $^\circ$) & (mJy beam$^{-1}$) \\
         \hline
         1 & BC188B & 2009 May 01 & BR \ FD \ HN \ LA \ MK \ NL \ OV \ PT \ SC & 1.20 $\times$ 0.35, --11 & 18.0 \\ 
         2 & BC188C & 2009 Oct. 19 & FD \ HN \ KP \ LA \ MK \ NL \ OV \ PT  & 2.60 $\times$ 0.30, --18 & 17.6 \\
         3 & BC188D & 2009 Nov. 09 & BR \ FD \ HN \ KP \ LA \ MK \ NL \ OV \ PT \ SC & 1.35 $\times$ 0.32, --15 & 14.9 \\
         4 & BC188E & 2010 May 01 & BR \ FD \ HN \ KP \ LA \ MK \ NL \ OV \ PT \ SC & 1.46 $\times$ 0.32, --14 & 17.9 \\
         \hline
         \end{tabular}
\end{center}
\tablefoot{Antenna codes are BR: Brewster, WA; FD:
Fort Davis, TX; HN: Hancock, NH; KP: Kitt Peak, AZ; LA: Los Alamos, NM; MK:
Mauna Kea, HI; NL: North Liberty, IA; OV: Owens Valley, CA; PT: Pie Town,
NM; and SC: Saint Croix, VI.
The restoring beams size and position angle (east of north) as well as the rms noise level at 22 GHz are listed in Columns 3 and 4.  }
   \end{table*}

Our observations were carried out using the 
NRAO\footnote{The National Radio Astronomy Observatory is a facility of the National 
Science Foundation operated under cooperative agreement by Associated Universities, Inc.}  
Very Long Baseline Array (VLBA) under program BC188. 
Since astrometric accuracy is improved by using background sources near in angle to the target, 
we conducted a brief survey on 2009 February 1 (BC188A) in order to find suitable phase-reference 
sources before starting parallax measurements. 
We observed 6 candidate background sources from the VLBA Calibrator Survey (VCS) 
with angular separations $<3.5$ degrees of OH\,231.8+4.2 at 22 and 43 GHz. 
The absolute position of the observed sources from the ICRF2 catalog and our measured brightnesses 
are listed in Table~\ref{table:survey}.

Our VLBA observations of H$_2$O and $\varv$ = 2, $J$ = 1 -- 0 SiO masers 
were performed over 8 hour tracks on 2009 May 01, Oct 19, Nov 09 and 2010 May 01. 
Observational parameters are listed in Table~\ref{table:obs}. The rest frequencies of the observed H$_2$O and SiO lines are 22235.08 and 42820.586 MHz, respectively. 
The dates were designed to optimize the parallax measurement by sampling near the maximum and 
minimum of the parallax signatures in Right Ascension.
We did not optimize for Declination, since its 
parallax signature is smaller than that for Right Ascension, and 
Declination position measurements generally have larger uncertainties 
for low declination sources such as our target. 

In order to measure a trigonometric parallax, we used phase-referencing observations, 
fast switching between the maser target and an extragalactic continuum source.
We used one background source (J0746--1555) for the H$_2$O maser 
and two background sources (J0746--1555 and J0748--1639) for the SiO maser observations. J0746-1555 is located 1.55$^\circ$ from OH\,231.8+4.2. The second calibrator at 43 GHz was added as a back up calibrator in the case that the primariy calibrator was not bright enough at the higher frequency. While J0748-1639 was brighter, it was located 2.4$^\circ$ from OH\,231.8+4.2, making it less suitable for astrometry.
We alternated between one block at 22 GHz and two blocks at 43 GHz. 
We switched sources every 40 seconds at 22 GHz and every 25 to 30 seconds at 43 GHz.
A strong source, J0530+1331, was observed approximately hourly to monitor delay and 
electronic phase differences among the IF bands.
In order to calibrate atmospheric delays for each antenna, 
we placed geodetic blocks \citep{Brunthaler2005,Reid09a} before the start, in the middle, 
and after the end of phase-referencing observations.  

The data were correlated in two passes with the VLBA DiFX 
software correlator\footnote{This work made use of the Swinburne University of Technology software
correlator, developed as part of the Australian Major National Research
Facilities Programme and operated under licence.} \citep{Deller}
in Socorro, NM. 
The four dual-polarized frequency bands of 8 MHz bandwidth were first processed 
with 16 spectral channels for each frequency band.  Next,
the frequency band containing maser emission was re-correlated with 1024 channels, 
giving channel spacings of 0.10 km s$^{-1}$ at 22 GHz and 0.05 km s$^{-1}$ at 43 GHz. 
The data reduction was performed with the NRAO's Astronomical Image Processing System (AIPS) 
package, using ParselTongue \citep{Kettenis} scripts developed for the Bar and Spiral 
Structure Legacy (BeSSeL) Survey\footnote{http://bessel.vlbi-astrometry.org} \citep{Brunthaler11} following the procedures described in \cite{Reid09a}.

\section{Results}

   \begin{figure}
   \centering
   \includegraphics[width=8cm]{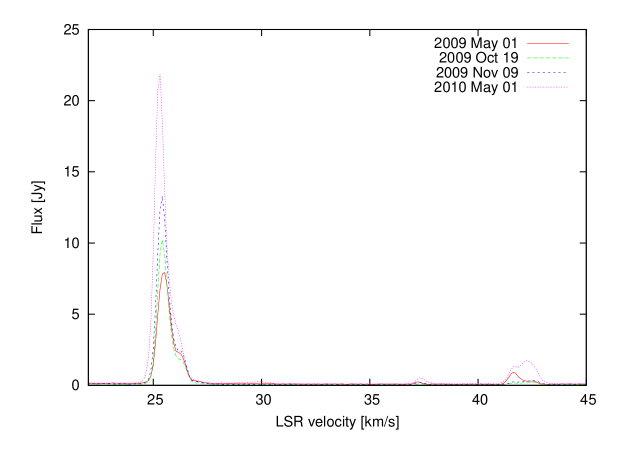}
      \caption{Spectra of the H$_2$O masers toward OH\,231.8+4.2 produced 
                  by scalar averaging the data over all time and baseline. 
				  Four different lines represent different epochs.}
         \label{FigSpectra}
   \end{figure}

\begin{figure*}
\centering
\resizebox{180mm}{!}{\includegraphics[]{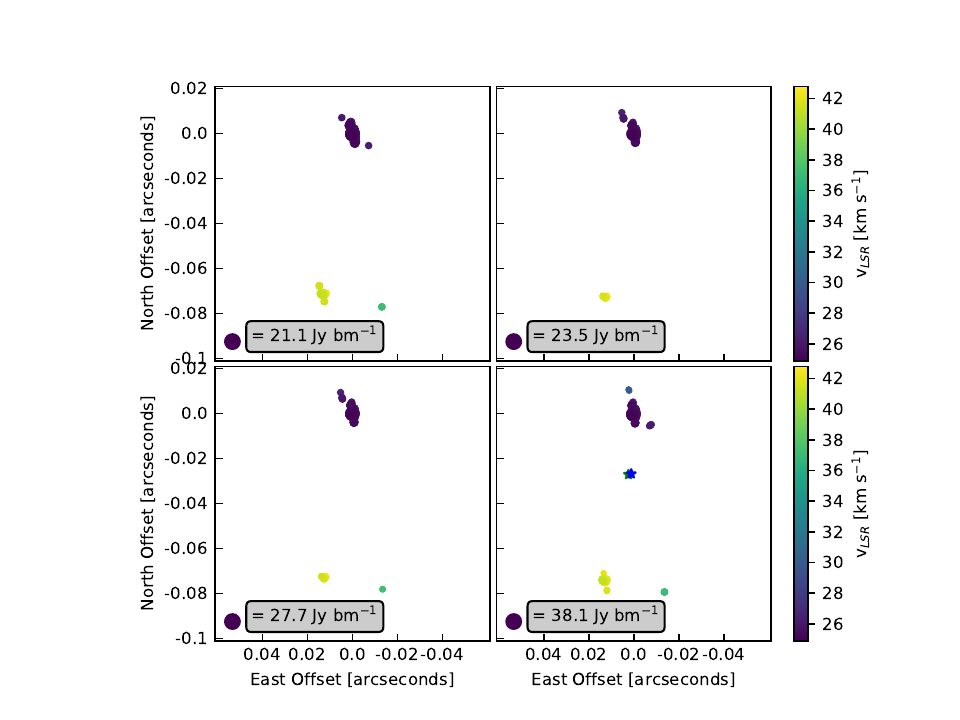}}
 \caption{Spatial distribution of the H$_2$O masers in OH\,231.8+4.2 at all epochs (top left: 1; top right: 2; bottom left: 3; bottom right: 4). 
The LSR velocities of the maser spots are indicated by the color bar to the right. The reference spot at $V_{\rm LSR}$ = 25.52 km s$^{-1}$ is located at (0,0). The blue and green stars in the bottom right map show the location of the SiO maser spots at 27.26 and 32.23 km s$^{-1}$, respectively.} 
   \label{FigMap}
\end{figure*}
   
\subsection{Spectrum and Spatial Distribution of Water Masers}

Figure~\ref{FigSpectra} shows the H$_2$O maser spectra toward OH\,231.8+4.2 at all four epochs, produced by scalar averaging the data over all observing time and baselines.  The H$_2$O masers span LSR velocities from 24 to 43 km s$^{-1}$, covering the systemic velocity of the nebula of 34 km s$^{-1}$ \citep{Alcolea01}. The maser features divide into two groups: blue-shifted features at $V_{\rm LSR}$ between 24 and 30 km s$^{-1}$ and red-shifted features between 37 and 43 km s$^{-1}$. Though the masers vary in flux density, the overall spectral appearance is qualitatively similar at all epochs, suggesting that many maser spots survive over the observing period of 1 year. We selected the strong maser spot at $V_{\rm LSR}$ of 25.52 km s$^{-1}$ as a reference spot. 

We defined a maser feature, when maser spots are detected stronger than 7$\sigma$ in 3 consecutive channels. The positions and intensities of the maser features are obtained from the peak emission from the epoch at which it was first detected during our observations. These are listed in Table~\ref{table:maser}. 
The spatial distributions of the H$_2$O masers in OH\,231.8+4.2 
relative to the reference maser spot at all epochs are shown in Figure~\ref{FigMap}. The H$_2$O masers are located within 40 mas in the east-west and 100 mas in the north-south directions. The blue-shifted components are located in the north, while the red-shifted components are in the south. The northern components are stronger than the southern components. The overall appearance is very similar between the epochs, with two and four maser spots newly appearing at the 2nd and 4th epoch, respectively.

\begin{table}
\caption{H$_2$O maser features toward OH\,231.8+4.2.}             
\label{table:maser}      
\centering          
\begin{tabular}{c c r r r c}     
\hline\hline       
ID  & $V_{LSR}$ &     S$_{peak}$  &     R.A.  &  Dec.  & Detected \\  
    & (km/s)   & (Jy beam$^{-1}$) &    (mas)  &  (mas) & epochs \\  
 \hline
\ 1 & 23.94 & 0.29 & 3.27 & --1.32 & - - - 4 \\
\ 2 & 24.57 & 0.28 & 5.18 & --2.02 & 1 2 3 4 \\
\ 3 & 25.52 & 21.12 &  0.0 & 0.0  & 1 2 3 4 \\
\ 4 & 26.05 & 9.21  & --0.18 & 1.27  & 1 2 3 4 \\
\ 5 & 26.05 & 2.11  & 4.52 & 6.69 & - 2 3 4 \\
\ 6 & 26.26 & 0.66  & --2.76 & --12.80  & 1 2 3 4 \\
\ 7 & 26.36 & 1.55  & --7.17 & --5.31  & 1 2 3 4 \\  
\ 8 & 26.78 & 0.53  & --10.60 & --1.11 & - - - 4 \\ 
\ 9 & 27.00 & 1.30  & 7.47 & 13.74  & 1 2 3 4 \\
10 & 27.21 & 0.91 & --1.38 & 3.01  & 1 2 3 4 \\
11 & 28.47 & 0.34 & --0.29 & 5.72  & 1 2 3 4 \\
12 & 28.89 & 0.42 & 0.17 & 6.66  & 1 2 3 4 \\
13 & 29.10 & 0.44 & 0.31 & 7.44  & 1 2 3 4 \\
14 & 30.05 & 0.50 & 2.25 & 10.36  & 1 2 3 4 \\
15 & 37.21 & 1.72 & --13.06 & --77.01 & 1 2 3 4 \\
16 & 38.27 & 0.25 & --31.41 & --56.38 & - - - 4 \\
17 & 40.27 & 0.33 & --16.80 & --77.07 & - - - 4 \\
18 & 41.64 & 5.27 & 13.62 & --71.14  & 1 2 3 4 \\
19 & 42.38 & 1.77 & 12.21 & --72.64 & - 2 3 4 \\
20 & 42.48 & 2.65 & 12.10 & --71.10  & 1 2 3 4 \\
\hline               
\end{tabular}
\end{table}

We detected 20 H$_2$O maser features in at least one epoch; this compares to 16 features detected by \cite{Desmurs}, 30 features by \cite{Leal-Ferreira}, and 20 features detected by \cite{Dodson2018}. Compared with these other VLBI observations of the H$_2$O masers, the overall structure of the features' distribution is similar, with blue-shifted features in the north and red-shifted features in the south. However, details of the spatial distribution are found to have changed over time. For example, we could not detect some northeast maser features at $V_{\rm LSR}$ $\sim$ 29 km s$^{-1}$ in \cite{Desmurs} and \cite{Leal-Ferreira}, but new maser features in the southern complex appeared at $V_{\rm LSR}$ between 38 and 41 km s$^{-1}$ in our observations.
The masers in the north have been consistently brighter by a factor of 3$-$4 in our obsersavtion as well as the previous observations from \cite{Desmurs} and \cite{Leal-Ferreira}. However, this has changed in the later observation by \cite{Dodson2018} where the southern masers are slightly brighter than the nothern masers.

\begin{figure*}[t]
\begin{center}
\includegraphics[width=13cm]{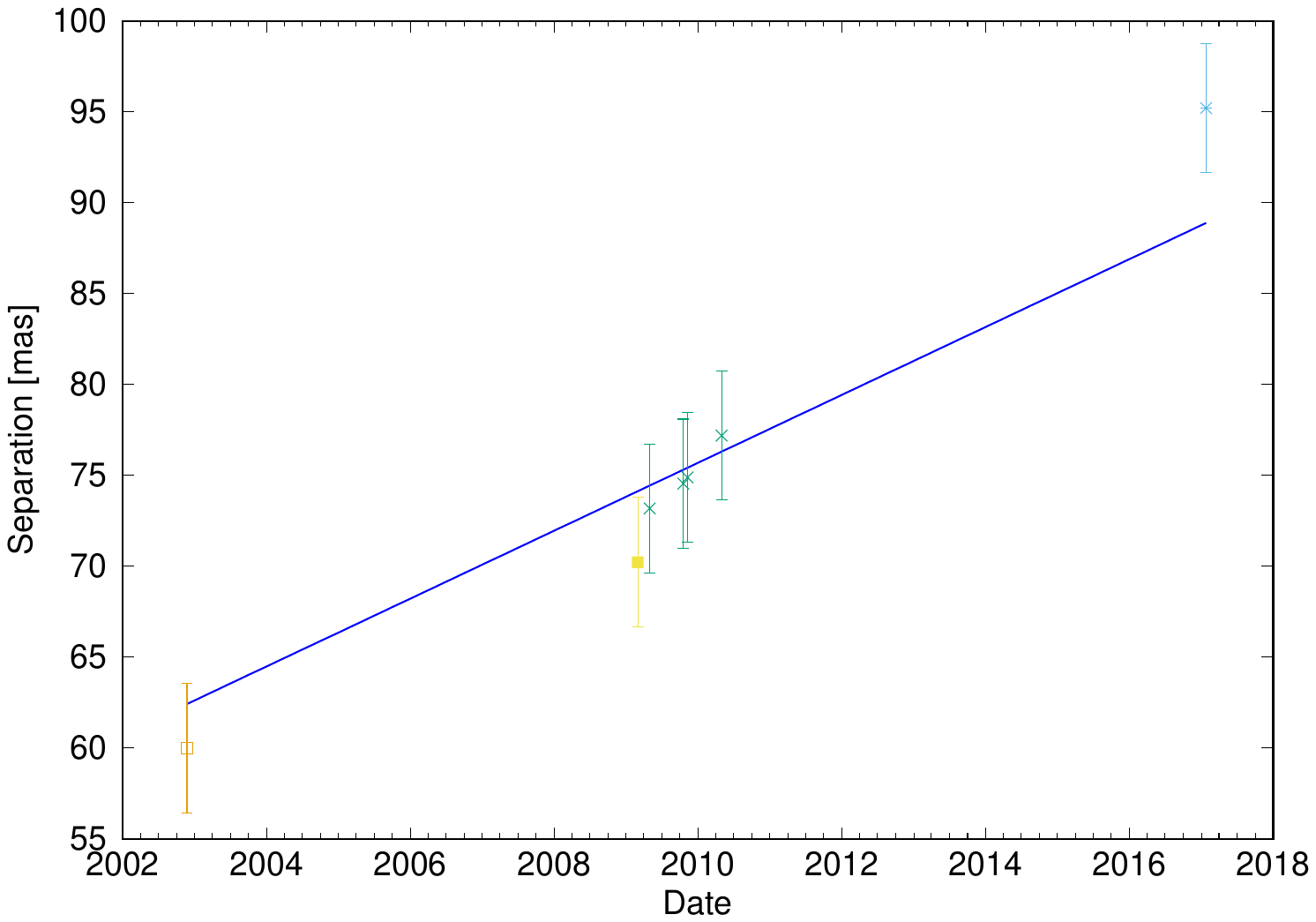} 
\caption{The median separation between the northern and southern water maser cluster in the observations from \cite{Desmurs} (open square), \cite{Leal-Ferreira} (filled square), \cite{Dodson2018} (blue x) and our own data (green x). The blue line is a linear fit to all data.}
\label{Fig:expand}
\end{center}
\end{figure*}

\cite{Dodson2018} combined a new observation in 2017 with the previously published observations to measure the median distance between the northern and southern clusters. We repeat this analysis by adding the data from our four epochs, and the results are shown in Fig.~\ref{Fig:expand}. A linear fit to all data, shows that the separation between the two clusters grows with 2.5 $\pm$ 0.2 mas yr$^{-1}$, corresponding to 17.3 $\pm$ 1.5 km s$^{-1}$ for our measured distance (see next section), consistent with the estimate of 19 km s$^{-1}$ in \cite{Dodson2018}. The uncertainty is estimated from the scatter in the individual measurements. The median velocity of the features in the northern and southern clusters are 26.3 and 40.8 km s$^{-1}$, respectively. Comparing the velocity difference in the plane of the sky (17.3 km s$^{-1}$ with the velocity difference in radial velocities (14.5 km s$^{-1}$), yields an inclination angle of $\sim$ 40$^\circ$, which is comparable to previous estimates of 36$^\circ$ \citep{Kastner92}. This expansion is much slower than the velocities seen on larger scales from the CO emission, which reach deprojected velocities of -210 km s$^{-1}$ and 376 km s$^{-1}$ for the northern and southern lobes, respectively \citep{Alcolea01}. This striking difference in expansion velocities between the water maser emission and the large scale CO outflow could be explained by a fast moving jet or outflow where the maser emission is exited by a fast outflow which interacts with the denser circumstellar environment, as proposed by \cite{TW86}.

\subsection{Parallax and Proper Motion of the Water Masers}

\begin{table*}
\begin{center}
\caption{Parallax and proper motions fits of the H$_2$O masers.}
\begin{tabular}{cccc} 
\hline \hline
LSR velocity & parallax  & $\mu_{x}$ & $\mu_{y}$ \\ 
(km s$^{-1}$) & (mas)  & (mas yr$^{-1}$) & (mas yr$^{-1}$)  \\ \hline
25.52 & 0.65 $\pm$ 0.01 & -4.35 $\pm$ 0.02 &  \ 0.71 $\pm$ 0.59 \\
37.32 & 0.65 $\pm$ 0.02 & -4.94 $\pm$ 0.07 & -1.65 $\pm$ 0.02 \\ 
41.85 & 0.66 $\pm$ 0.03 & -4.05 $\pm$ 0.11 & -1.75 $\pm$ 0.07 \\  \hline
\end{tabular} \\
\label{table:para}
\end{center}
\end{table*}

\begin{figure*}
\centering
   \includegraphics[width=13cm]{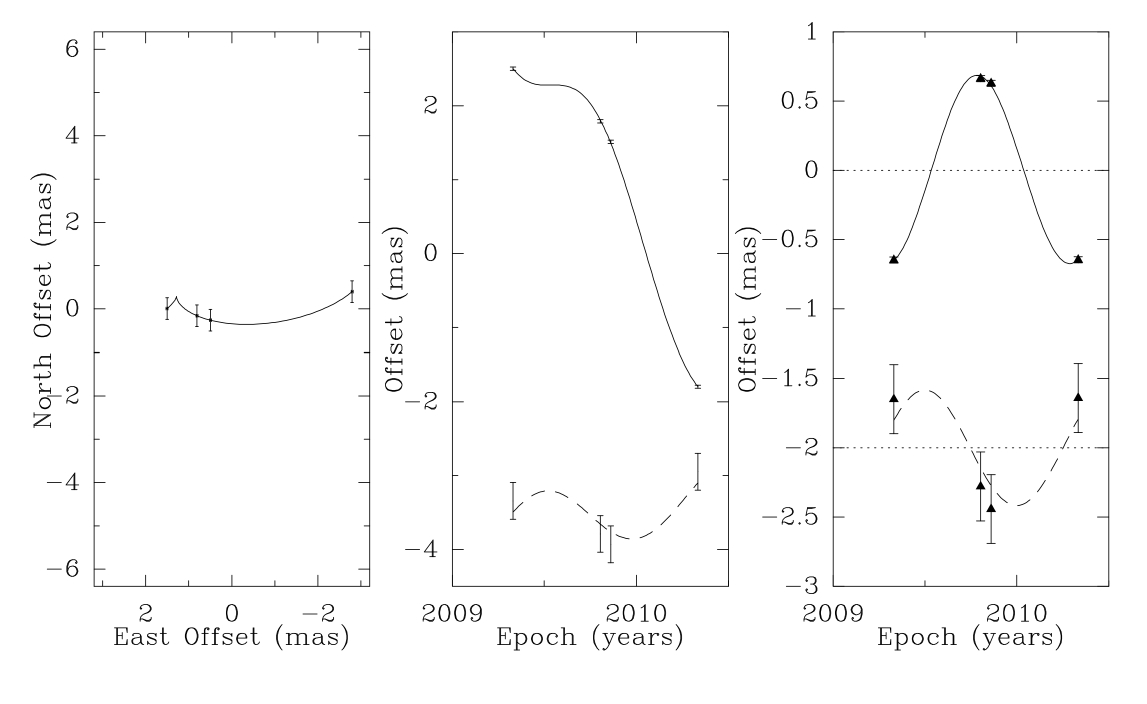}
 \caption{Parallax and proper motion fit for OH\,231.8+4.2. 
For each plot, points with error bars are position measurements of the H$_2$O maser spot at $V_{\rm LSR}$ 
of 25.52 km s$^{-1}$ relative to the background quasar J0746--1555. 
Left panel: motion of the maser on the sky.
Middle panel: the change of position eastward (solid line) and northward (dashed line) as a function of time.  The northward data have been shifted downward for clarity.
Right panel: same as the middle panel after removing the best-fitting linear proper motion. 
} 
   \label{FigPara}
\end{figure*}

Figure~\ref{FigPara} shows position measurements of the H$_2$O maser spot at a LSR velocity of 25.52 km s$^{-1}$ (the reference spot) relative to the background continuum source J0746--1555 
over a timespan of 1 year. The position offsets are with respect to $\alpha$(J2000) = $07^{\rm h}42^{\rm m}16^{\rm s}.9175$ and $\delta$(J2000) = $-14^\circ42' 50{"}.0344$. 

We obtained an annual parallax and proper motion by variance-weighted least-squares fitting the position versus time data for the H$_2$O maser spots in OH\,231.8+4.2 detected at all four epochs. While the parallaxes should be identical within measurement uncertainties for all spots, the proper motions could reflect differing internal motions of the spots. Our results are listed in Table~\ref{table:para}. The measured parallaxes are in good agreement, suggesting any unresolved structural changes in the maser spots are minimal. The estimated parallax is 0.65 $\pm$ 0.01 mas (statistical uncertainty). 

Our observations employed a single background quasar and had only 4-epochs. Since the east-west data dominate the parallax solution, there are only 4 data points to solve for 3 parameters (parallax, position offsets, and motion). This leaves only one degree of freedom to estimate the full uncertainty, including systematic effects for the parallax. While this does not bias the parallax estimate itself, the uncertainty can be over- or underestimated by $\sim$44\% as shown in \cite{Reid17}. Therefore, we conservatively estimate a systematic uncertainty, predominantly owing to uncompensated tropospheric delays, of $\pm$0.02 mas.  Thus, our final result is 0.65 $\pm$ 0.01 (statistical) $\pm$ 0.02 (systematic) mas. This corresponds to a distance of 1.54$\pm$ 0.05 kpc. In order to get an estimate of the absolute motion of the AGB star, we averaged the two red-shifted features and then split the difference between the the red and blue values. This results in $\mu_x$ = --4.45 $\pm$ 0.45 mas yr$^{-1}$ eastward and $\mu_y$ = --0.50 $\pm$ 1.2 mas yr$^{-1}$ northward.

\subsection{SiO Maser Emission}
Unlike the H$_2$O maser emission, the SiO maser emission was very weak during our observations.  We were able to detect only two maser features at LSR velocities of 27.26 and 32.23 km s$^{-1}$ phase-referenced to the quasar J0746--1555 at the fourth epoch (BC188E). These two SiO maser spots are separated by 1.3 mas and lie between the two clusters of H$_2$O masers, as shown in the last panel of Fig.~\ref{FigMap}, presumably indicating the location of the central AGB star. \cite{Dodson2018} used the Korean VLBI Network and its unique capability to observe the water and SiO masers simultaneously to do an accurate astrometric alignment between the two maser species. They also find that the SiO maser emission is located between the two clusters of water maser emission. We note that both observations locate the SiO masers slightly closer to the northern cluster than to the southern cluster, with a distance ratio approximately at 40/60. 

The absolute position of the spots at 27.26 km s$^{-1}$ in the fourth epoch is $\alpha$(J2000.0) = $07^{\rm h}42^{\rm m}16^{\rm s}.91726$ and $\delta$(J2000.0) = $-14^\circ42' 50".0607$.  The uncertainty in the absolute position is dominated by the uncertainty in the position of the calibrator, which is $\pm$0.72 mas in right ascension and $\pm$1.32 mas in declination \citep{VCS2}. \cite{Desmurs} observed the SiO masers with the VLBA in phase reference mode in 2003. They also identified two maser spots at ~27 and 32 km s$^{-1}$ and obtained an absolute position for the maser spot at 27 km s$^{-1}$ which differs by 38.9 mas in right ascension and 3.3 mas in declination from our positions. Since the 2003 observation was in early July, while our 2010 epoch was conducted in early May, the contribution of the parallax effect is insignificant, and the difference in position can be attributed to the proper motion of the central Mira star QX\,Pup. This gives a proper motion of -5.7 $\pm$ 0.2 mas yr$^{-1}$ in right ascension and -0.48 $\pm$ 0.2 mas yr$^{-1}$ in declination.

This proper motion is similar to the proper motion estimated from the water maser emission. While the motion in declination agrees perfectly, there is a 2.5 sigma difference in the proper motion in right ascension. Since the SiO maser emission originates much closer to the star itself, than the water masers, the SiO maser proper motion is a much better representation of the motion of the star and will be adopted here.

\section{Gaia EDR3 data}
OH\,231.8+4.2 is located only $\sim$11' toward the north-east of the center of the open cluster M\,46 (NGC\,2437). The cluster has an r$_{50}$ radius (i.e. the radius containing 50\% of the stars) of 10.5' \citep{Hunt2023}. Therefore, OH\,231.8+4.2 is believed to be associated with M\,46. The distance of M\,46 can be also estimated with Gaia EDR3 data. To select members of the cluster, we first selected all stars with a distance of less than 0.8 degrees from the cluster center. We used only sources where the ratio of the parallax and the formal parallax uncertainty was larger than 10. To reduce the contamination of possible binary systems we follow the recommended cuts from \cite{Fabricius21}:
\begin{itemize}
    \item \texttt{ruwe} < 1.4
    \item \texttt{ipd\_frac\_multi\_peak} $\leq$ 2 
    \item \texttt{ipd\_gof\_harmonic\_amplitude} < 0.1
\end{itemize}

Next, we used only stars that had proper motions within $\pm1$ mas yr$^{-1}$ of the cluster proper motion as given in \cite{Babusiaux18}. We then reduced the allowed proper motion range, by calculating the standard deviation of the proper motions, and allowing only stars that have proper motions within 3$\sigma$ of the cluster mean proper motion. We repeated this process until the standard deviation of the proper motions did not change anymore. This resulted in the final cuts of $|\texttt{pmra}+3.86|<0.4$ mas yr$^{-1}$ and $|\texttt{pmdec}-0.4|<0.4$ mas yr$^{-1}$. These cuts gave us a list of 1316 possible cluster members.

\begin{figure}[t]
\begin{center}
\includegraphics[width=8cm]{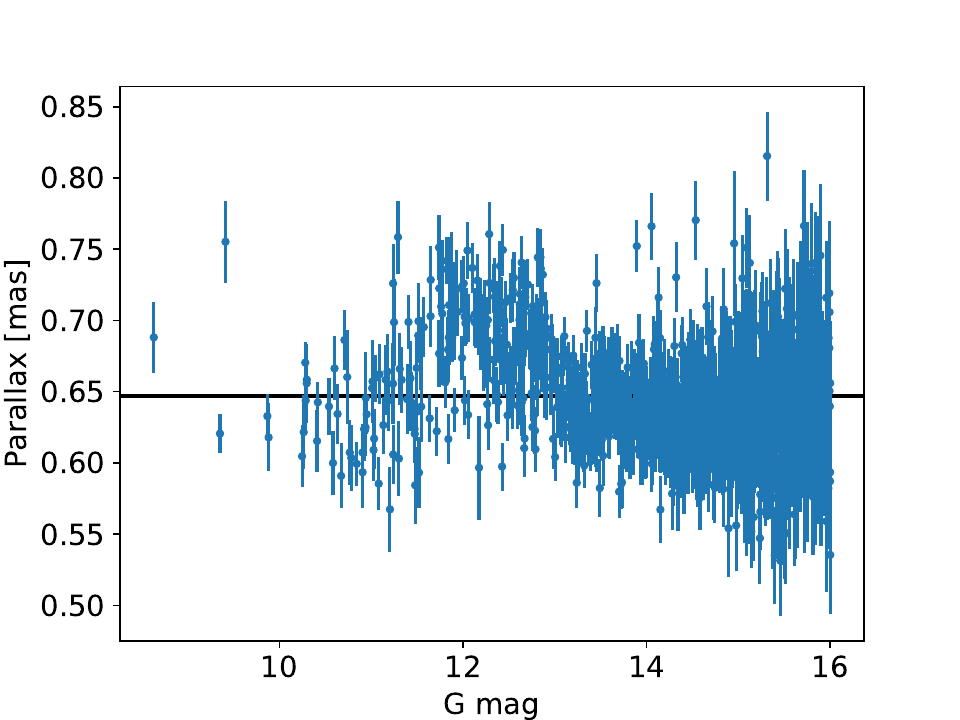} 
\includegraphics[width=8cm]{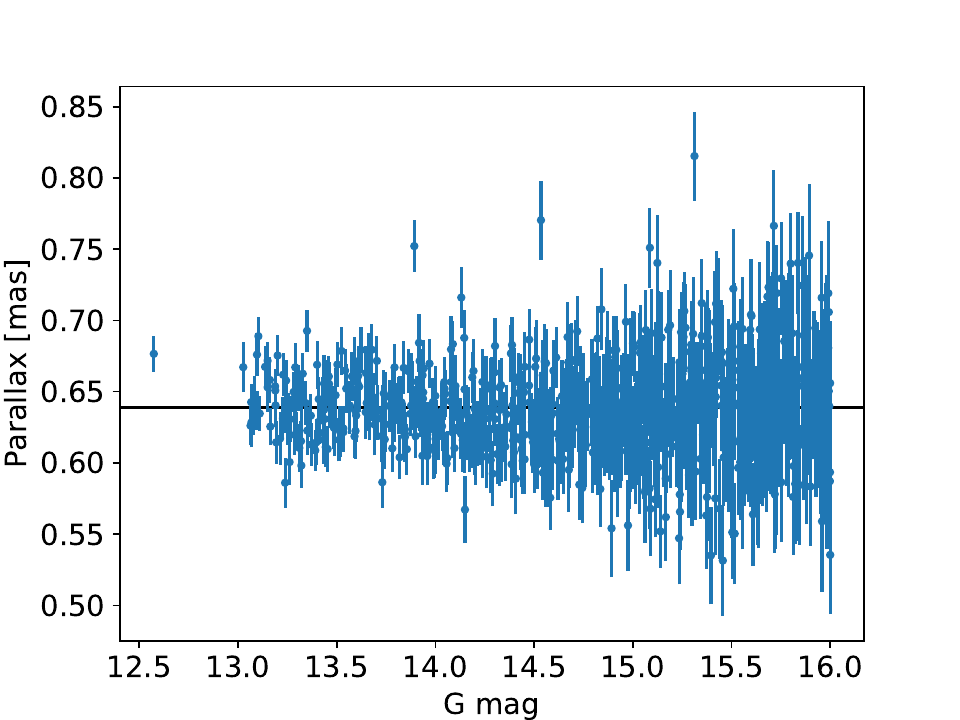} 
\caption{{\bf Upper panel:} Parallaxes of individual stars in M\,46 plotted against the apparent G magnitude of the star. The horizontal line marks the average parallax. {\bf Lower panel:} Same as the upper panel, but removing stars with \texttt{astrometric\_excess\_noise\_sig}>2. The average parallax changed by 8 $\mu$as.}
\label{Fig:gaia_all}
\end{center}
\end{figure}

We then apply the parallax zero-point correction from 
\citep{Lindegren21b}. While the validity of this zero-point correction has been confirmed by a number of studies, \cite{Huang21} show, that there is a small overcorrection in the magnitude range from 13<G<16. Hence, we apply a additional correction of $5 (G - 13)$ $\mu$as for this magnitude range as described in \cite{Leung22}, and remove any stars which are fainter than G>16 mag. There are several stars, which are clearly fore- or background objects, so we also remove stars with parallaxes which differ from the mean parallax by more than 5 times the parallax uncertainty of the star. These additional cuts leave 1036 stars. To check the consistency of the parallax measurements, we plot the parallax value versus the apparent G magnitude of the stars in Fig.~\ref{Fig:gaia_all} (upper panel). Here, we can clearly see, that many stars within the G magnitude range from about 11.5 < G < 12.5 have on average larger parallaxes. It is not clear what causes this apparent bias. One possibility is that the zero-point correction is not adequate for these stars. However, all of these sources have a non-zero \texttt{astrometric\_excess\_noise}. We therefore added one additional cut and removed all stars with \texttt{astrometric\_excess\_noise\_sig}$>2$ (see section 5.1.2 of \cite{Lindgren2012}). This results in our final list of 747 stars which for a much cleaner sample as shown in the lower panel of Fig.~\ref{Fig:gaia_all}.

Using this clean sample of member stars, the average parallax of all stars is 0.639 mas (and a median of 0.638 mas) with an formal error of only 1.2 $\mu$as.  This formal error underestimates the true uncertainty, since the measurements for the 747 stars are not truly independent. For example, the overcorrection seen by \cite{Huang21} has not been independently confirmed and there are spatial correlations on small scales \citep{Lindegren21a} which introduce an additional bias (Note that about 50\% of all stars are within 0.2$^\circ$ of the cluster center). Using Equation 24 from \citet{Lindegren21a}, we therefore assign a systematic uncertainty of 10 $\mu$as to the result, which gives us a final parallax of 0.639$\pm$0.001$\pm$0.010 mas.

The average proper motion of the final 747 stars is -3.861 mas yr$^-1$ in Right Ascension and 0.399 mas yr$^{-1}$ in declination. The formal uncertainties are 0.005 mas yr$^{-1}$ in both directions. The velocity dispersion is 0.13 mas yr$^{-1}$ in both directions, which corresponds to 0.9 km s$^{-1}$. While there is also a bias in the proper motions and \cite{CGB21} provide a recipe for a correction, these corrections are only defined for stars with a G magnitude brighter than 13. However, these stars are all except one removed by our cut on the astrometric excess noise. However, if we apply the corrections to the 289 stars brighter than G$<$13 from our sample before this cut, the bias corrections range form -0.013 to +0.017 mas yr$^{-1}$ with a median of 0.012 mas yr$^{-1}$ in Right Ascension, and from 0.025 to 0.061 mas yr$^{-1}$ with a median of 0.051 mas yr$^{-1}$ in Declination. Since it is not clear how this bias affects the weaker stars in our final sample, we therefore assume additional systematic uncertainties of 0.012 mas yr$^{-1}$ in Right Ascension and 0.051 mas yr$^{-1}$ in Declination. This leads us to the final proper motions of -3.861 $\pm$ 0.005 $\pm$ 0.012 mas yr$^{-1}$ and 0.399 $\pm$ 0.005 $\pm$ 0.051 mas yr$^{-1}$ in Right Ascension and Declination, respectively.

\section{Discussion}

The agreement between our VLBI maser parallax for OH\,231.8+4.2 of 0.65 $\pm$ 0.01 (statistical) $\pm$ 0.02 (systematic) mas and the Gaia EDR3 parallax for M\,46 of 0.639 $\pm$ 0.001 (statistical) $\pm$ 0.01 (systematic) mas within the uncertainties clearly shows that the nebula is part of the open cluster. 

Previously, the distance to OH\,231.8+4.2 has been estimated to be 1.3$\pm$0.4 kpc \citep{Kastner92}, assuming that OH\,231.8+4.2 is a member of the cluster \citep{Jura}. On the other hand, the kinematic distance using the systemic LSR velocity of 34 \kms\, is $\approx3.7$ kpc, more than twice our measured distance. Our parallax distance of OH\,231.8+4.2 is 1.54 kpc, with 5\% error, which suggests that the kinematic distance is anomalous.  

In previous studies, the luminosity was estimated using the distance of  1.3$\pm$0.4 kpc. While this distance is consistent with our result within their quited uncertainty, it is 16\% smaller than our result.   Since the luminosity depends on a square of distance, luminosity estimates quoted by simply assuming a distance often without uncertainty were underestimated by 34\%. With the parallax distance, the luminosity of OH\,231.8+4.2 is 1.4 $\times$ 10$^4$ L$_{\odot}$, with about 10\% uncertainty from the distance. 

\begin{table}
\begin{center}
\caption{Proper motions measurements of OH\,231.8+4.2 and M\,46}
\begin{tabular}{ccc} 
\hline \hline
Method &  $\mu_{x}$ & $\mu_{y}$ \\ 
&  (mas yr$^{-1}$) & (mas yr$^{-1}$)  \\ \hline
H$_2$O  & -4.45 $\pm$ 0.45 & -0.50 $\pm$ 1.20 \\
SiO  & -5.7 $\pm$ 0.2 & -0.48 $\pm$ 0.2 \\ 
Gaia & -3.861 $\pm$ 0.012 & 0.399 $\pm$ 0.05 \\  \hline
\end{tabular} \\
\label{table:para}
\end{center}
\end{table} 

\begin{figure}[t]
\begin{center}
\includegraphics[width=8cm]{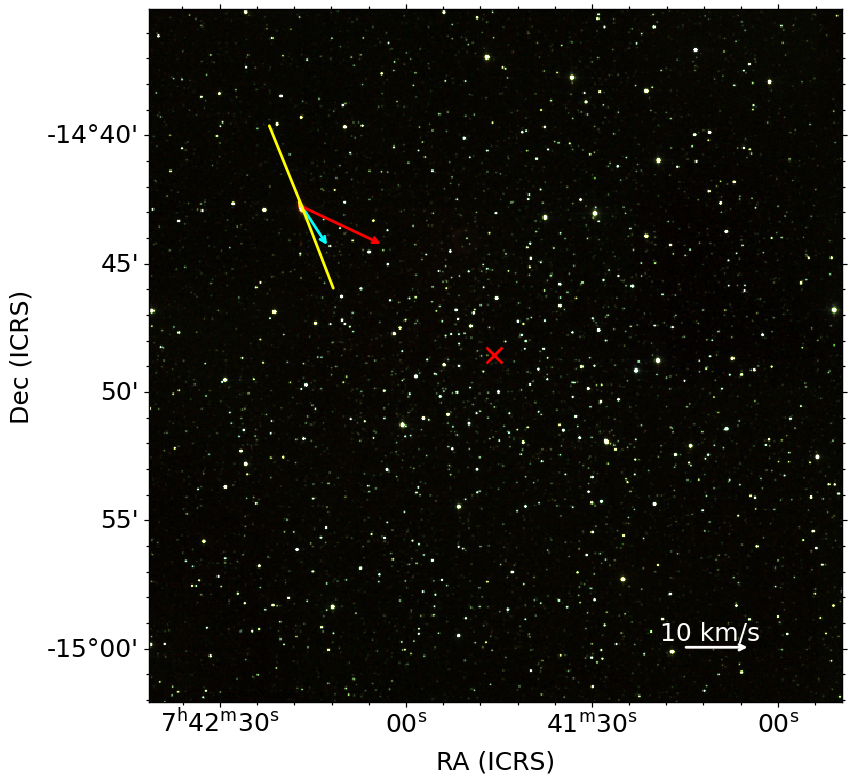} 
\caption{2MASS color composite of M\,46. The center of the open cluster is marked with an red X. OH\,231.8+4.2 is also seen and its tangential motions derived from the water masers (cyan) and SiO masers (red) relative to M\,46 are also shown. The yellow line marks the orientation of the large scale CO outflow.  The white arrow in the bottom right corner corresponds to a tangential motion of 10 \kms.}
\label{Fig:2Mass}
\end{center}
\end{figure}

While the parallax between our VLBI measurements and the estimate from Gaia is in excellent agreement, we find a significant difference in the measured proper motions of the nebula and the stars in the cluster. Fig.~\ref{Fig:2Mass} shows a 2MASS color composite image of the open cluster M\,46 and the location of OH\,231.8+4.2. Also shown are the proper motions relative to the cluster as determined from the water and SiO masers, and the direction of the large scale CO outflow.
If we assume that the motions of the SiO masers represent the motion of the AGB star, it with moves -1.84 mas yr$^{-1}$ in Right Ascension and -0.88 mas yr$^{-1}$ in Declination relative to the cluster of stars. This would correspond to tangential motions of 12.8 km s$^{-1}$ and 6.1 km s$^{-1}$ at our measured distance and is significantly larger than the velocity dispersion of the cluster itself. This apparent motion is directed at a position angle of 245$^\circ$ East of North.
From the water masers, we would obtain tangential motions of 4.0 km s$^{-1}$ in Right Ascension and 6.1 km s$^{-1}$ in Declination and a position angle of 213$^\circ$ East of North. In any case, the motion is directed roughly toward the center of the cluster.


The heliocentric radial velocity of the cluster is 46.9$\pm$1 km s$^{-1}$ \citep{Frinchaboy2008, Kiss2008}, corresponding to an LSR velocity of 28.9 km s$^{-1}$, while the systemic LSR velocity of the nebula is 34 km s$^{-1}$ \citep{Alcolea01}. Here, we also find a difference of approximately 5 km s$^{-1}$.

\subsection{A stellar merger scenario could explain the proper motion conundrum}
As discussed above, the proper motion of \oh does not agree with that of the cluster M\,46.
Ejected stars from open clusters are known for decades as runaway stars \citep{Blaauw1961} with velocities higher than 30 km~s$^{-1}$ or walkaway stars with velocities of a few to tens of km~s$^{-1}$ \citep{deMink2014}. Two mechanisms have been proposed to explain the large velocities of the stars:  dynamical ejection caused by few-body interactions \citep{Poveda1967, Lennon18, Kalari19} and supernova explosions in a binary system \citep{Blaauw1961, Renzo2019, Chrimes23}.

\oh is located just outside the r$_{50}$ radius of the cluster, and it is moving currently torward the center of M\,46. Therefore, the system formed even further outside, where the stellar density is much lower than in the dense core of the cluster. Furthermore, the fraction of ejected stars drops with decreasing mass of the stars \citep[e.g. ][]{Oh2016}. Hence, it seems unlikely that the discrepant velocity is caused by a dynamical interaction with other stars. 

In the binary supernova scenario, the system formed as a triple system, where the most massive star exploded in a supernova. \cite{Renzo2019} showed that the large majority of stars ejected by supernova explosions in binary systems have velocities smaller than 30 km~s$^{-1}$, consistent with the motion of \oh. However, it is not clear whether the remaining two stars would have remained bound after such a disruptive event.

While both explanations can not be ruled out and need further investigation with detailed simulations, they do not explain the origin of the bipolar outflow seen in the source. 

Another possible origin for  the velocity discrepancy and the bipolar outflow could be that the phenomena observed in \oh are the result of a stellar merger,  as suggested by \cite{Staff2016}. 
A merger can be the result of orbital decay in a multiple stellar system. Since the velocities of the remnant objects are determined by the dynamics of the merger process, they are decoupled from original space velocity of the system. This is well-illustrated by the example of Source I and other sources in the Orion Kleinmann Low nebula within the core of Orion Molecular Cloud 1, whose proper motions have been measured with the Very Large Array 
(sources BN and n) \citep{Gomez2008}. These authors show that the three sources move away from each other with on-the-sky velocities of 15 to 26 \kms\ and shared a common origin $~500$~y ago. The aftermath of the 
explosion accompanying the merger is observed in vibrationally excited H$_2$ emission and CO emission which shows a spherically symmetric distribution. That Orion I is indeed the 
result of a merger had been proposed by \citep{BallyZinnecker2005}. 
At radio and mm wavelengths, Source I has the morphology of an edge-on disk. The disk is associated with strong SiO maser emission which expands away from it that is at the base of large scale bipolar outlow traced by thermal SiO emission. 

Recently, a small sample of sources have been closely studied that, 
like Orion I, share characteristics with OH\,231.8+4.2, in particular are they characterised by high velocity and chemically peculiar molecular material. These are so-called ``red novae'', due to their sudden appearance in the sky, which is accompanied by copious production of dust (and molecules). 
Discussing the prominent case of the red nova (RN) V838 Mon, which experienced a major outburst in early
2002, \citet{SokerTylenda2003} proposed the merger of two main sequence stars as the energy source. The main compact remnant of the merger is an M-type stellar object with red supergiant characteristics which is associated with SiO maser emission \citep{Deguchi2005, Ortiz2020}.
SiO maser emission also is an outstanding characteristic of Orion I \citep[see][and references therein]{MentenReid1995}.

\citet{Kaminski2018} present millimeter/submillimeter-wavelength observations with the Atacama Large Millimeter Array (ALMA) and the Submillimeter Array (SMA) of three well-studied Galactic red novae, V4332 Sgr, V1309 Sco, and V838 Mon. Although they only covered a limited frequency range, they detected emission from CO, SiO, SO, 
SO$_2$ in all three sources. Additionally, a line from H$_2$S was covered and detected toward V 838 Mon. These sulphur-bearing species have motivated  the Rotten Egg Nebula's name! \citet{Kaminski2015,Kaminski2017} presented a much larger amount of molecular line data for the archetypal red nova CK Vul which was discovered as Nova Vul 1670 more than 350 y ago.

While the molecular line emission of all of the mentioned RNe is characterised by wide line widths of up to 600 km s$^{-1}$ which exceed OH\,231.8+4.2's $\sim 200$~km s$^{-1}$ , only toward CK Vul an extended well-developed bipolar outflow has been discovered. We note that both for OH\,231.8+4.2 and  CK Vul shocked ionized material has been found at in their outflows \citep{Reipurth1987,Hajduk2007}

We note that the RN events for V838 Mon and V1309 Sco occurred in 2002 and 2008, respectively, too short a time for an angularly extended outflow to develop, given these objects distances, which have been estimated as 3.7 and 5.6 kpc, respectively \citep{Tylenda2011, Ortiz2020}. 
In contrast, the CK Vul merger happened in 1670, while a dynamical age of 800 years has been estimated for the CO outflow in OH\,231.8+4.2 \citep{Sanchez2022}.

To summarize: Many of the characteristics of the  emission observed toward OH\,231.8+4.2  closely resemble what is found toward  RNe. 
Apart from the bipolarity and their content in sulphur-bearing molecules, OH\,231.8+4.2 and the merger sources Orion I and CK Vul share other chemical characteristics. Toward CK Vul, several metal-bearing molecules were discovered, namely NaCl and KCl which had been first discovered toward the high mass-loss carbon-rich AGB star IRC+10216, but not in other AGB objects. Recent sensitive ALMA observation have found these species in a compact circumbinary disk in \oh \citet{Sanchez}
and in the disk in the  Orion I \citep{Ginsburg2019}.
Taken together, the discrepant velocity and it chemical peculiarities are {supporting the hypothesis of \cite{Staff2016} that  OH\,2381.8+4.2 is a result of a stellar merger.

\section{Conclusion}
This study presents a precise measurement of the distance to the protoplanetary nebula OH\,231.8+4.2, also known as the Rotten Egg or Calabash Nebula, using VLBI observations of H$_2$O and SiO masers, yielding a trigonometric parallax of 0.65 $\pm$ 0.01 (stat) $\pm$ 0.02 (syst) mas, corresponding to a distance of 1.54 $\pm$ 0.05 kpc. This result agrees remarkably well with the Gaia EDR3-derived distance to the nearby open cluster M\,46, confirming a physical association between the nebula and the cluster. However, the nebula exhibits a significant proper motion and radial velocity offset relative to M\,46, suggesting it may be a dynamically disturbed object. Plausible explanations for these anomalies are a past stellar merger, dynamical ejections, or a supernova explosion. Since the system formed most likely in the outskirts of the cluster, the dynamical ejection scenario seem unlikely. And detailed modelling would be required to investigate whether the current binary system could have survived a supernova explosion of a former third star in the system. Although these other scenarios for the measured velocity difference can not be ruled out based on the current data, a previously proposed stellar merger origin could explain the velocity offset, the chemical peculiarities, and the bipolar outflow seen in the system.

\begin{acknowledgements}
This research made use of hips2fits,\footnote{https://alasky.cds.unistra.fr/hips-image-services/hips2fits} a service provided by CDS. We thank the anonymous referee for helpful suggestions which improded this manuscript.
\end{acknowledgements}

%
%

\end{document}